# Propagating Conditions and the Time of ICMEs Arrival:
# A Comparison of the Effective Acceleration Model with ENLIL and DBEM Models

Evangelos Paouris[1,2], Jasa Calogovic[3], Mateja Dumbovic[3], M. Leila Mays[4], Angelos Vourlidas[1,5], Athanasios Papaioannou[1], Anastasios Anastasiadis[1], Georgios Balasis[1]

*1: Institute for Astronomy, Astrophysics, Space Applications & Remote Sensing of the National Observatory of Athens (IAASARS-NOA), Penteli, GR 15236, Greece.*

*2: Faculty of Physics, National and Kapodistrian University of Athens, Athens, GR 15784, Greece.*

*3: Hvar Observatory, Faculty of Geodesy, University of Zagreb, Kaciceva 26, HR-10000, Zagreb, Croatia.*

*4: NASA Goddard Space Flight Center, Greenbelt, MD 20771, USA.*

*5: The Johns Hopkins University Applied Physics Laboratory, Laurel, MD 20723, USA*

**Abstract:** The Effective Acceleration Model (EAM) predicts the Time-of-Arrival (ToA) of the Coronal Mass Ejection (CME) driven shock and the average speed within the sheath at 1 AU. The model is based on the assumption that the ambient solar wind interacts with the interplanetary CME (ICME) resulting in constant acceleration or deceleration. The upgraded version of the model (EAMv3), presented here, incorporates two basic improvements: (a) a new technique for the calculation of the acceleration (or deceleration) of the ICME from the Sun to 1 AU and (b) a correction for the CME plane-of-sky speed. A validation of the upgraded EAM model is performed via comparisons to predictions from the ensemble version of the Drag-Based model (DBEM) and the WSA-ENLIL+Cone ensemble model. A common sample of 16 CMEs/ICMEs, in 2013-2014, is used for the comparison. Basic performance metrics such as the mean absolute error (MAE), mean error (ME) and root mean squared error (RMSE) between observed and predicted values of ToA are presented. MAE for EAM model was 8.7±1.6 hours while for DBEM and ENLIL was 14.3±2.2 and 12.8±1.7 hours, respectively. ME for EAM was -1.4±2.7 hours in contrast with -9.7±3.4 and -6.1±3.3 hours from DBEM and ENLIL. We also study the hypothesis of stronger deceleration in the interplanetary (IP) space utilizing the EAMv3 and DBEM models. In particularly, the DBEM model perform better when a greater value of drag parameter, of order of a factor of 3, is used in contrast to previous studies. EAMv3 model shows a deceleration of ICMEs at greater distances, with a mean value of 0.72 AU.

## 1. Introduction

Interplanetary coronal mass ejections (ICMEs) are the main sources of the most intense variability of the interplanetary space environment (Gosling 1993; Zhang et al., 2007; Paouris and Mavromichalaki, 2017a; Papaioannou et al., 2020). Their corresponding geospace disturbances affect a wide range of technological systems in space (satellites, communications, GPS systems) and on ground (ground induced currents on power grids and pipelines), as well as human health (astronauts in orbit, and crew and passengers on high latitude flights; Gopalswamy, 2009; Hanslmeier, 2010 and references therein). To safeguard against those hazards, we need better predictions of the ICME properties upon Earth arrival. Vourlidas, Patsourakos and Savani, (2019) reviewed the open issues related to the CME forecasting, as well as the current capabilities of the scientific community to address them. The pertinent open issues concerning CME/ICME forecasting are: (a) the Time-of-Arrival (ToA) at Earth, (b) the speed of the ICME at 1 AU or speed-of-arrival (SoA) and (c) the magnitude and the sign of the southward component (Bz) of the ICME magnetic field. The latter is the most important parameter for the prediction of the strength of the resulting geomagnetic storm (see e.g. Gosling et al., 1993).

In this work, we focus on the ToA problem. In recent years, many models have been developed to predict the ToA. These models are very diverse also in complexity and the physical approaches as well as in the number of assumptions taken into account. The different categories according to Vourlidas, Patsourakos and Savani, (2019) are: (a) empirical models (Gopalswamy et al., 2001; Schwenn et al., 2005; Kilpua et al., 2012; Colaninno et al., 2013; Mostl et al., 2014; Vrsnak et al., 2014; Makela et al., 2016; Rollett et al., 2016; Wood et al., 2017; Mostl et al., 2017; Paouris and Mavromichalaki 2017a; 2017b), (b) drag-based models (Vrsnak et al., 2013; Shi et al., 2015; Hess and Zhang, 2015; Dumbovic et al., 2018; Napoletano et al., 2018; Kay et al., 2020), (c) physics-based shock models (Corona-Romero et al., 2015; 2017), (d) MHD models (Millward et al., 2013; Shiota and Kataoka, 2016; Riley et al., 2018; Wold et al., 2018; Pomoell and Poedts, 2018) and, (e) machine learning models (Sudar et al., 2016; Liu et al., 2018).

Previous studies mention various techniques to predict the ToA with interesting results. However, there is a major issue related to the validation and verification of these different models. The researchers are performing verification and validation studies in such a way that the comparison between the models is nearly impossible. In particular, they are using different samples of CME/ICME events, input CME parameters and different sets of metrics as well as they are using different set of events to train or tune their models (see e.g. Fry et al., 2001; Gopalswamy et al., 2005; McKenna-Lawlor et al., 2006; Taktakishvili et al., 2009; Vrsnak et al., 2014; Mays et al., 2015; Paouris & Mavromichalaki, 2017b; Dumbovic et al., 2018; Wold et al., 2018). One common statistical metric used in the evaluation of model performance is the mean absolute error (MAE) between the predicted and the actual ToA. The MAE ranges between 3.5 and 17.7 hours for various models (Vourlidas, Patsourakos and Savani, 2019). The most critical aspect for proper model performance comparison is the event sample. In particular, the MAE of 3.5 hours arising from samples of 7 (Hess and Zhang, 2015) and 8 (Corona-Romero et al., 2015) events, while a worse MAE of 17.7 hours arose from a much larger sample of 214 ICMEs (Paouris and Mavromichalaki, 2017b). However, Wold et al. (2018) studied a large sample of 273 events and they found a MAE of almost 10.4 hours. A possible explanation of the variability of MAE results in these works is that MAE used primarily as a measure of how the study was set up (if the models were actually optimized or not for the examined sample) and secondarily for the accuracy of the model. The mean value of MAE for all of the models is 10±2 hours. This value reflects the state-of-the-art that can be achieved with the currently available data (Vourlidas, Patsourakos and Savani, 2019).

The Community Coordinated Modeling Center (CCMC) and in particular the CME Arrival Time and Impact Working Team (https://ccmc.gsfc.nasa.gov/assessment/topics/helio-cme-arrival.php) is leading an effort to quantify and establish a set of metrics together with a CME validation event set which will provide a benchmark against current and future models which can be assessed (see Verbeke et al. 2018). The most efficient and reliable method to compare the results from the various models is the development of a common set of events and a common set of metrics.

In our study we focus on the improved version of Effective Acceleration Model (hereafter, EAMv3) and the comparison of this version with the ensemble version of the drag-based model (DBEM) and WSA-ENLIL+Cone ensemble model. In order to have a reliable comparison of EAMv3 with other models we employ the sample used by Mays et al. (2015) and Dumbovic et al. (2018), focusing only on the 16 CMEs/ICMEs which finally arrived on Earth. This is the first time that three different models are tested with the same set of ICMEs thus providing a consistent way to compare their metrics. These metrics are the mean absolute error (MAE), the mean error (ME) and the root mean squared error (RMSE).

The paper is structured as follows. In Section 2, we explain the upgrades of EAMv3 model and in Section 3, we are presenting the validation of EAMv3 model through the comparison with WSA-ENLIL+Cone and DBEM models. In Section 4, we focus on the analysis concerning the overestimation of ToA from all models studying the possible stronger deceleration of the ICMEs in the interplanetary (IP) space, and finally the conclusions presented in the last section (Section 5).

## 2. Effective Acceleration Model
### 2.1 Effective Acceleration

Paouris and Mavromichalaki (2017a, hereafter Paper I), using a large sample of 266 well-established CME/ICME pairs, estimated the acceleration or the deceleration of the ICME in the interplanetary space. To calculate the acceleration (or deceleration) simple kinematic equations and input parameters were used. These parameters were the CME onset time, transit time, initial and 1 AU speeds and the distance (1 AU).

The fundamental assumption of the EAM model is that the ICME interaction with the ambient solar wind results in a constant "effective acceleration" of the CME from Sun to 1 AU. The core of the model is an empirical relation for the acceleration as a function of the initial speed of the CME, which is obtained from coronagraph data. From the derived value of the effective acceleration, it is possible to estimate the ToA of the CME or of the preceding shock, via simple kinematic equations.

In Paouris and Mavromichalaki (2017b, hereafter Paper II) the effective acceleration was calculated for each CME/ICME pair using the relation:

$$\alpha = \frac{\Delta \upsilon}{\mathrm{TT}} = \frac{\upsilon_f - u_0}{\mathrm{TT}} \ [\mathrm{m\ s^{-2}}] \qquad (1)$$

where $\upsilon_f$ [km s$^{-1}$] is the mean solar wind speed in the sheath (the area between the shock driven by the ICME and the main part of the ICME) using data from Solar Wind Electron Proton Alpha Monitor (SWEPAM; McComas et al., 1998) of Advanced Composition Explorer (ACE; Stone et al., 1998), and $u_0$ [km s$^{-1}$] is the initial speed of the CME obtained by Large Angle and Spectrometric Coronagraph (LASCO; Brueckner et al., 1995) coronagraphs of Solar and Heliospheric Observatory (SOHO; Domingo et al. 1995). This speed ($u_0$) is the linear speed and not the speed in the direction of propagation and is affected by the projection effects (see e.g. Burkepile et al., 2004; Vrsnak et al., 2007; Paouris et al., 2020 and references therein). The transit time of ICME *TT* [s] is calculated using the times of first appearance of CME in LASCO coronagraphs and the arrival time of the shock on ACE. The acceleration (or deceleration) was calculated only for ICMEs which were associated with a shock, i.e. 214 events from Paper I list, from 1996 up to 2009.

The previous version of our model (EAMv2) was relying on a second order polynomial fit between the linear speed $u_0$ from LASCO coronagraphs and the effective acceleration α as follows:

$$\alpha = 1.41392 \left[ \text{ms}^{-2} \right] - 26.30 \times 10^{-4} \left[ 10^{-3} \text{s}^{-1} \right] u_0 - 1.14717 \times 10^{-6} \left[ 10^{-6} \text{m}^{-1} \right] u_0^2 \quad (2)$$

with a very high correlation coefficient of *r* = 0.981 when excluding outliers. We consider as outliers, the points outside the 99% prediction bands. In Equation 1 we see that the effective acceleration and the linear speed of CME are not independent.

Nevertheless, the uncertainties in Equation 1 originate from both speeds $v_f$ and $u_0$. Both CME speed and angular width of the CME from coronagraph data are subject to projection effects. CMEs are observed in coronagraph images due to the Thompson scattered electrons. This emission is optically thin and CMEs which are actually 3D structures are represented as a 2D structures in the coronagraph field-of-view (FOV) (see e.g. Burkepile et al., 2004; Vrsnak et al., 2007; Paouris et al., 2020 and references therein).

To minimize these uncertainties, we use a new approach to calculate the effective acceleration, eliminating the speed of the flow in the sheath $v_f$ from Equation 1. In particular, we use:

$$\alpha = \frac{2d}{\text{TT}^2} - \frac{2u_0}{\text{TT}} \quad [\text{m s}^{-2}] \quad (3)$$

where $u_0$ is the projected linear speed from LASCO, *TT* is the transit time between the first appearance of the CME in the LASCO C2 and the arrival time of the shock at 1 AU from ACE data, and d is the Sun-Earth distance taken to be 1 AU.

These new values of effective acceleration as a function of the linear speed of CME lead to a new 2$^{nd}$ order polynomial function of the form:

$$\alpha = 2.8476 \left[ \text{ms}^{-2} \right] - 23.70 \times 10^{-4} \left[ 10^{-3} \text{s}^{-1} \right] u_0 - 2.65072 \times 10^{-6} \left[ 10^{-6} \text{m}^{-1} \right] u_0^2 \quad (4)$$

with a correlation coefficient of *r* = 0.982.

The scatter plots between the linear speed of the CME from LASCO and the effective acceleration using Equation 1 and Equation 3 are presented in Figures 1a and 1b. In Figure 1a, the distribution of the points is narrower, especially for CME linear speeds less than 1000 km s$^{-1}$, in contrast to the distribution of the points in Figure 1b. The calculation of effective acceleration in Figure 1b, is based only on the linear speed of CME, without considering $v_f$ and its uncertainties, as the transit time is the same in both Equations 1 and 3. At this point, we should mention that in both Figures 1a and 1b we do not present the uncertainties but the calculated effective acceleration as a function of linear CME speed.

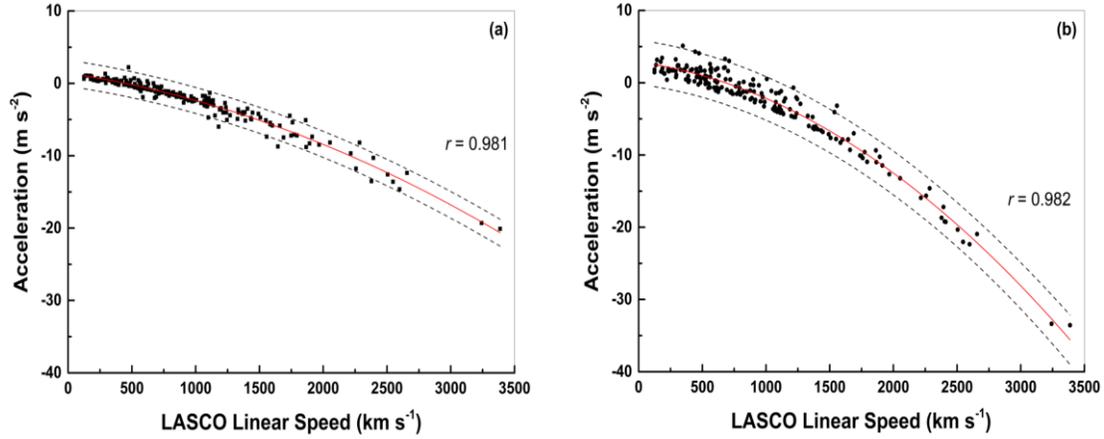

**Figure 1**: Effective acceleration in relation with the linear speed of CMEs using Equation 1 (Figure 1a) and Equation 3 (Figure 1b). The red line indicates the best 2nd order polynomial fit with correlation coefficients of 0.98 in both cases. The black dashed lines are indicating the 99% confidence intervals.

## 2.2 Correction of CME Linear Speed

In Paper II we used the Leblanc et al. (2001) empirical method to correct the linear speed of CMEs for projection effects. The method is based on the association of each CME to their source on the solar disk (e.g. solar flare) because the location of the flare associated CMEs is well identified contrary to CMEs associated with filaments. Coupling ICME to their sources was considered for the event list of Paper I. The requirements for Leblanc's method are the projected CME half-angular width, the linear speed, and the heliographic position (latitude and longitude) of the associated solar flare. Excluding CMEs associated with filaments as well as halo CMEs, where the true half angular width is highly uncertain, there were 87 CME/ICME events associated with solar flares. Then, for this sample, we applied Leblanc's method and finally an empirical relation between the linear speed $u_0$ of the CME (projected speed on the plane-of-sky POS from LASCO) and the radial speed $u_r$ (corrected speed) was founded:

$$u_r = 41.516 + 1.027 \cdot u_o \text{ [km s}^{-1}\text{]} \qquad (5)$$

This linear relation leads to a minor improvement of ToA estimation of about one hour in our model. As shown in Table 1, the initial MAE of 18.58 hours (EAMv1) decreased to 17.65 hours (EAMv2) but remained above the average of other models with mean MAE of almost 10 hours (Vourlidas, Patsourakos and Savani, 2019). We note that the majority of the models using > 40 events have average MAE of almost 13.5 hours, according to the Table 1 of Vourlidas, Patsourakos and Savani (2019), and in all cases the MAEs are greater than 10.4 hours. One explanation about the small MAE values in small samples is that in most of the cases the events are carefully picked and the models are actual fine-tuned for those events. In those cases, the reported MAE is more of a measure of the best case scenario for performance. Another factor is whether the MAE comes from an actual forecast or a hindcast, which is actually a process where the forecast performed using some parameters that were not available originally (e.g. multi-viewpoint observations to obtain the CME radial speed).

Based on this minor improvement derived from the utilization of the corrected speed (Equation 5), we search for a possible correction of the CME linear speed ($u_0$) to be applied before running the model. In particular, we use a linear equation of the form:

$$u_r = b_0 u_0 + b_1 \quad (6)$$

In particular, we reverse the problem by parameterizing the corrected speed of CMEs, $u_r$, in order to minimize the MAE of the whole sample. Then, the factors $b_0$ and $b_1$ were calculated using the sample of 214 events from Paper II yielding values of $b_0 = 0.543$ and $b_1 = +45.6$ with a minimum MAE value of 12.74 hours. This version of the model gives ToA within ±5 hours for 25% of the sample and within ±10 hours for 50% of the sample. Only 11 events (5% of the sample) have a ToA error larger than ±30 hours. The various EAM versions, are compared in Table 1, for the same 214 event sample.

**Table 1:** The metrics for the previous and the current version of EAM model

| Model Version | Description / Input parameters used | ME | MAE | RMSE |
|---|---|---|---|---|
| EAMv1 | • Acceleration from Equation 1 <br> • LASCO linear speed | -3.03±1.53 | 18.58±0.87 | 22.47±0.95 |
| EAMv2 | • Acceleration from Equation 3 <br> • deprojected CME speed using Equation 5 | -0.28±1.48 | 17.65±0.85 | 21.55±0.96 |
| EAMv3 | • Acceleration from Equation 3 <br> • Linear relation for CME speed using Equation 6 | +3.66±1.05 | 12.74±0.64 | 15.82±0.75 |

## 3. Validation of the EAM Upgraded Version

As mentioned earlier, the model developers are performing their own verification and validation analysis and it is almost impossible to compare the various results because the samples are all different. To avoid this, for validation of EAMv3 upgraded version we use a sample of 16 events used also by Mays et al (2015) and Dumbovic et al. (2018). We note that the sample of 214 events from Paper I covering the period January 1996 – December 2008, is independent from the sample of 16 events used here for the validation of the model which covers the period April 2013 – September 2014. The events and their parameters are shown in Table 2. In particular, Columns 2 and 3 give the CME onset date and time (in UT), and Column 4 provides the mean value of the CME speed from the ensemble modeling (Mays et al., 2015). We note that the mean CME speed is the average speed obtained from all the available runs for each one of the 16 CMEs. CME measurements and simulation summary results for each event utilized also by Mays et al. (2015) and Dumbovic et al. (2018) are available at https://iswa.ccmc.gsfc.nasa.gov/ENSEMBLE/. Columns 5 and 6 show the ICME arrival date and time (in UT) based on in situ observations at L1 from ACE. Finally, Columns 7-9 contain the prediction error between the predicted arrival time and the actual arrival time of the ICME at L1, given by the equation: $\Delta t_{err} = t_{predicted} - t_{observed}$, from ENLIL, DBEM and EAMv3 models, respectively. The latter three lines of Table 2 contain the statistical metrics of MAE, ME and RMSE for all three models utilized the sample of 16 common events. The standard errors are calculated with a simple bootstrap method with replacement which yields $10^6$ resamples.

**Table 2:** Details for the 16 common CMEs/ICMEs and the calculated metrics for ENLIL, DBEMv1 and EAMv3 models.

| Event | CME onset | | Mean Speed | ICME Arrival | | $\Delta t_{err}$ [h] | | |
|---|---|---|---|---|---|---|---|---|
| | Date [UT] | Time [UT] | [km s$^{-1}$] | Date [UT] | Time [UT] | ENLIL | DBEMv1 | EAMv3 |
| 1 | 11/04/2013 | 07:24 | 1026.9 | 13/04/2013 | 22:13 | -16.0 | -14.6 | +00.15 |
| 2 | 21/06/2013 | 03:12 | 2008.0 | 23/06/2013 | 03:51 | -14.8 | -13.9 | +00.80 |
| 3 | 30/08/2013 | 02:48 | 883.7 | 02/09/2013 | 01:56 | -17.4 | -07.8 | -05.13 |
| 4 | 29/09/2013 | 20:40 | 1005.7 | 02/10/2013 | 01:15 | +02.9 | -01.8 | +10.68 |
| 5 | 06/10/2013 | 14:39 | 774.5 | 08/10/2013 | 19:40 | +26.5 | +05.3 | +14.55 |
| 6 | 07/01/2014 | 18:24 | 2424.8 | 09/01/2014 | 19:39 | -19.4 | -23.0 | -07.13 |
| 7 | 30/01/2014 | 16:24 | 880.2 | 02/02/2014 | 23:20 | -13.2 | -19.9 | -14.22 |
| 8 | 12/02/2014 | 05:39 | 737.1 | 15/02/2014 | 12:46 | -13.0 | -20.9 | -11.45 |
| 9 | 18/02/2014 | 01:25 | 881.7 | 20/02/2014 | 02:42 | +13.8 | +10.2 | +15.57 |
| 10 | 19/02/2014 | 16:00 | 882.4 | 23/02/2014 | 06:09 | -17.8 | -33.1 | -21.30 |
| 11 | 25/02/2014 | 01:09 | 1393.8 | 27/02/2014 | 15:50 | -17.6 | -24.5 | -06.60 |
| 12 | 23/03/2014 | 03:36 | 720.5 | 25/03/2014 | 19:10 | +05.8 | +08.6 | +04.93 |
| 13 | 02/04/2014 | 13:36 | 1538.5 | 05/04/2014 | 10:00 | -15.0 | -24.9 | -13.88 |
| 14 | 18/04/2014 | 13:09 | 1404.2 | 20/04/2014 | 10:20 | -05.2 | -07.4 | +11.63 |
| 15 | 04/06/2014 | 15:48 | 585.8 | 07/06/2014 | 16:12 | +04.7 | +06.6 | -00.05 |
| 16 | 19/06/2014 | 17:12 | 564.5 | 22/06/2014 | 18:28 | -02.3 | +06.4 | -00.35 |
| | | | | | MAE | 12.84±1.65 | 14.31±2.18 | 08.65±1.57 |
| | | | | | ME | -06.13±3.27 | -09.67±3.43 | -01.36±2.66 |
| | | | | | RMSE | 14.53±1.52 | 16.93±2.25 | 10.69±1.42 |

In Figure 2 the histogram of the CME arrival time prediction error ($\Delta t$) for each one of the 16 events is presented. It is evident that all models show a tendency for early prediction and underestimate the ToA, i.e. the ICME arrived later than the predicted time and as a result the arrival time error, $\Delta t$ is negative. In particular, the mean error (ME) for ENLIL, DBEM and EAMv3 models is equal to -9.67, -6.13 and -1.36 hours respectively. In 10 cases (62.5%) all models converge with the same results for the error $\Delta t$, specifically in three cases all models are overestimating ($\Delta t > 0$) the ToA, and in seven cases are underestimating ($\Delta t < 0$) the ToA.

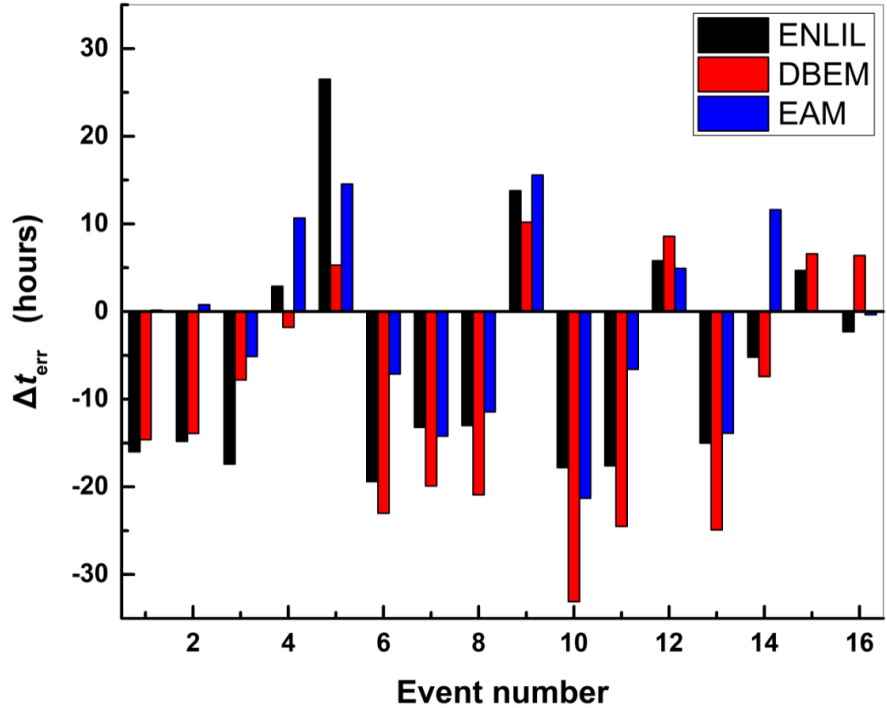

**Figure 2**: Errors (Δ*t*) between the predicted and the actual arrival time of the ICME for each case of 16 events for ENLIL, DBEM and EAM models

## 4. Effects of CME Propagating Conditions on ToA

The WSA-ENLIL+Cone ensemble model and DBEM have ME of -6.13 and -9.67 hours, respectively. The negative values, imply that the predicted ToA precedes the actual ToA. For our sample of 16 ICMEs all three models underestimate the ToA; however, the EAMv3 model shows a smaller ME of only -1.36 hours.

The improvement on the metrics of EAMv3 model, as it was described in Section 2, is based on the new approach for the calculation of the effective acceleration as well as the linear relation for the CME speed. This linear relation has a slope of almost 0.5 indicating that the speeds which serves as input in EAMv3, should be much smaller than the ones calculated using coronagraph white light images. The speeds, used by Mays et al. (2015) and Dumbovic et al. (2018), are the 3D speeds obtained by the triangulation method utilizing data from Solar TErrestrial RElations Observatory (STEREO; Kaiser et al., 2008) A and B spacecrafts. In particular, the CME initial parameters are determined using the Stereoscopic CME Analysis Tool (StereoCAT) developed by CCMC (see Mays et al., 2015). StereoCAT tracks specific CME features, based on triangulation of transient CME features manually identified using two different coronagraph fields of view. From this process the 3D speed, position, and the projected angular width are derived.

The question which arises here is the following: if we assume that the 3D speeds are very close to the real ones, then it is possible to have a stronger deceleration, perhaps in distances beyond 30 solar radii which is close to the limit of the FOV of the coronagraphs? Especially, for the very fast CMEs only a few points are available in the height-time measurements for the calculation of CME kinematics and as a result it is probable to lose this information, i.e. of deceleration of a CME in larger distances.

What does this value (0.5) of the slope possibly indicates? A potential answer might be that the deceleration of CMEs in the IP space due to drag is more important and it is insufficiently captured from the models. To investigate this scenario, we perform an analysis with EAMv3 and DBEM. The ToA problem is reversed for these 16 common events. In particular, we utilized the known values of onset and arrival CME times at Earth to find the optimal values of input parameters enforcing $\Delta t = 0$ between the predicted by the model and the actual ToA. For the EAMv3 model, the CME speed and acceleration (or deceleration) were fine-tuned to get $\Delta t = 0$. We note that we are not making any speed corrections using Equation 6 for the EAMv3 reverse modelling. In the first part of our analysis we calculate the optimal CME initial speed ($u_0$) to take $t_{predicted} = t_{observed}$ ($\Delta t = 0$) applying Equation 4. The optimal CME speed and the corresponding acceleration from Equation 4 are presented in columns 3 and 4 of Table 3. The second part utilizes the mean CME speed (column 5 in Table 3) assuming that this CME speed is the "true" initial speed and then we calculate the optimal effective acceleration (column 6 in Table 3) to take $\Delta t = 0$. For comparison, we present the LASCO $2^{nd}$ order speed at a distance of 20 solar radii in column 2 of Table 3.

The DBEM model uses more input parameters, such as the drag parameter (*gamma* - $\gamma$), the solar wind speed (*w*), and the CME kinematic and geometric properties (onset date & time, onset radial distance, initial speed, angular half-width and longitude). For each event, DBEM reverse modeling was applied using the latest version of DBEM (i.e. v25) to find optimized $\gamma$ and *w*. The CME properties which were used as an input were taken by Dumbovic et al. (2018). In DBEM after the user provides input values and their uncertainties, the ensemble is generated randomly assuming that each input parameter follows a normal distribution with observational input value as mean and standard deviation derived from 3σ uncertainty. For that purpose, the median values of the CME parameters from the ensembles in Mays et al. (2015) and Dumbovic et al. (2018) were calculated for each CME event. For each event, we perform reverse modelling with DBEM using the whole parameter space of possible *w* and $\gamma$ values, searching for [*w*, $\gamma$] combinations which result in both perfect ToA and SoA. Perfect ToA match is defined within ±1 hour of actual observed ToA, and perfect SoA match is defined ±10 km s$^{-1}$ of actual observed SoA. However, DBEM reverse modelling does not result in unique [*w*, $\gamma$] solution and mathematically it is possible that solutions are obtained outside of realistically expected *w* and $\gamma$ ranges (as e.g. derived in Vrsnak et al., 2013). Therefore, we constrain parameter space of possible *w* and $\gamma$ values to only "realistic" values of [300, 600] km s$^{-1}$ and [0.01, 0.59] × 10$^{-7}$ km$^{-1}$, respectively. These ranges were obtained for this particular sample by searching best fit $\gamma$ values for perfect ToA and SoA separately, using measured values of *w* upstream of the shock/sheath and behind the ICME trailing edge for each event. The measured *w* values were obtained from in-situ data from the Solar Wind Experiment (SWE; Ogilvie et al., 1995) on the Wind spacecraft (Szabo, 2015). We note that we assume that the parameter space of possible *w* and $\gamma$ values are the same for each event.

From 50 000 DBEM runs performed for each event we obtained minimum, maximum and median values of $\gamma$ and *w* that are able with the other input parameters to produce the perfect ToA (Figure 3). The corresponding optimal median values of *w* and $\gamma$ are also presented in Table 4. We note that for one event (event 10 in Table 4) no optimized *w* and $\gamma$ could be obtained using the CME input and constrained parameter space of possible *w* and $\gamma$ values as described above. The possible reasons for this might be unrealistic CME input, unrealistic [*w*, $\gamma$] parameter space or impossibility of drag-based equation to describe the propagation of this particular CME, however, the deeper analysis is out of the scope of this paper.

From this analysis, two interesting conclusions can be drawn. First, the DBEM underestimation of the ToA is related to the underestimated drag. The optimal input values for the drag coefficient $\gamma$ are, in all cases, higher than the values used in Dumbovic et al. (2018). Second, the input CME speeds seem to be overestimated if we compare the speeds in columns 3 and 5 of Table 3. In particular, 14 cases (87.5%), has optimal CME speeds lower than the mean values calculated using coronagraph white light images. We have an indication here that a possible correction might be needed not only for POS speeds (e.g. LASCO linear speed) but also for measurements obtained from triangulation (CME input speeds in Mays et al. (2015) and Dumbovic et al. (2018)). We do not state that the speeds are wrong, but the CME propagation is insensitive to coronal speeds. The overestimation of the CME initial speed, especially for fast CMEs, is also mentioned in Dumbovic et al. (2018) and in Mays et al. (2015). Furthermore, the optimal effective acceleration is negative (column 6 in Table 3) in 14 cases which implies that the majority of the events decelerates with much lower values in contrast to the acceleration values of column 4 in Table 3, in order to achieve $\Delta t = 0$.

**Table 3:** Input parameters and their optimal values for EAMv3 model to take minimum error ($\Delta t = 0$) for the sample of 16 ICMEs.

| Event | LASCO speed at 20 $R_{Sun}$ | Optimal CME input Speed[a] | Acceleration[b] from Equation 4 | Mean CME input speed | Optimal acceleration[c] |
|---|---|---|---|---|---|
| | [km s$^{-1}$] | [km s$^{-1}$] | [m s$^{-2}$] | [km s$^{-1}$] | [m s$^{-2}$] |
| 1 | 819 | 607.4 | 0.4535 | 1026.9 | -3.2785 |
| 2 | 1903 | 1171.7 | -3.2650 | 2008.0 | -13.1164 |
| 3 | 884 | 387.2 | 0.8704 | 883.7 | -2.3441 |
| 4 | 1164[d] | 959.1 | 0.5173 | 1005.7 | -2.3541 |
| 5 | 822[e] | 937.2 | 1.1668 | 774.5 | 0.0050 |
| 6 | 1714 | 1044.8 | -5.3001 | 2424.8 | -18.0869 |
| 7 | 940 | 182.7 | 0.8802 | 880.2 | -2.5821 |
| 8 | 494[f] | 180.8 | 1.2640 | 737.1 | -1.5729 |
| 9 | 712[g] | 1056.2 | 0.8760 | 881.7 | -0.6436 |
| 10 | 510 | 49.3 | 0.8741 | 882.4 | -2.6473 |
| 11 | 2069[h] | 589.2 | -0.7610 | 1393.8 | -6.5986 |
| 12 | 834 | 574.3 | 1.3065 | 720.5 | -0.6642 |
| 13 | 1468 | 443.7 | -1.2978 | 1538.5 | -7.6167 |
| 14 | 1245[i] | 1286.8 | -0.7985 | 1404.2 | -6.0326 |
| 15 | 1035[j] | 362.3 | 1.6350 | 585.8 | -0.0728 |
| 16 | 451[k] | 343.3 | 1.6844 | 564.5 | 0.0455 |

a: Optimal CME input (launch) speeds $u_0$ for EAMv3 model which produce for each event $\Delta t = 0$.
b: For the speeds in (a) we calculate the acceleration from Equation 4.
c: The optimal acceleration values for $\Delta t = 0$ when utilizing the mean CME input speed as $u_0$.
d: The CME onset time in CDAW is 22:12
e: The CME onset time in CDAW is 14:43
f: The CME onset time in CDAW is 06:00
g: The CME onset time in CDAW is 01:36
h: The CME onset time in CDAW is 01:26
i: The CME onset time in CDAW is 13:26
j: The CME onset time in CDAW is 12:48
k: The CME onset time in CDAW is 17:00

**Table 4:** Input parameters and their optimal values for DBEM model to take minimum error ($\Delta t = 0$) for the sample of 16 ICMEs.

| Event | Drag parameter ($\gamma$)[a] | Optimal Drag par. ($\gamma$)[b] | Solar wind speed (w)[c] | Optimal solar wind speed (w)[d] |
|---|---|---|---|---|
| | [$10^{-7}$ km$^{-1}$] | [$10^{-7}$ km$^{-1}$] | [km s$^{-1}$] | [km s$^{-1}$] |
| 1 | 0.1 | 0.38 | 350 | 400 |
| 2 | 0.1 | 0.33 | 350 | 449 |
| 3 | 0.1 | 0.34 | 350 | 412 |
| 4 | 0.1 | 0.28 | 350 | 472 |
| 5 | 0.1 | 0.21 | 350 | 506 |
| 6 | 0.1 | 0.42 | 350 | 402 |
| 7 | 0.1 | 0.37 | 350 | 371 |
| 8 | 0.1 | 0.42 | 350 | 324 |
| 9 | 0.1 | 0.21 | 350 | 492 |
| 10 | 0.1 | -[e] | 350 | -[e] |
| 11 | 0.1 | 0.42 | 350 | 369 |
| 12 | 0.1 | 0.27 | 350 | 513 |
| 13 | 0.1 | 0.48 | 350 | 358 |
| 14 | 0.1 | 0.27 | 350 | 469 |
| 15 | 0.1 | 0.29 | 350 | 469 |
| 16 | 0.1 | 0.30 | 350 | 463 |

a: Drag parameter($\gamma$) values used in Dumbovic et al. (2018).
b: Optimal values of drag parameter ($\gamma$) in DBEM model for $\Delta t = 0$.
c: Solar wind speed (w) used in Dumbovic et al. (2018).
d: Optimal values of solar wind speed (w) in DBEM model for $\Delta t = 0$.
e: Optimal $\gamma$ and w could not be obtained using reverse DBEM modelling for this event

An important aspect of our analysis is the role of drag forces on the ToA (see the review by Manchester et al., 2017 and references therein). The best way to study our hypothesis, that the deceleration of CMEs in the interplanetary space due to drag forces is more important, especially for fast CMEs, was to use the drag-based ensemble model (DBEM). In all cases the optimal drag parameter ($\gamma$) is greater than 2.1-4.8 times in contrast to the $\gamma$ value which has been used in Dumbovic et al. (2018) (see Table 4). Similar results were also observed for the speed of the ambient solar wind (*w*) and for 14 cases *w* is greater than the value used in the same study. The $\gamma$ ranges for the CME sample are plotted in Figure 3a. The solar wind speed (*w*) ranges are shown in Figure 3b. The optimal $\gamma$ values are much higher than the values estimated previously (solid red line), which clearly indicates that the drag parameter was underestimated in Dumbovic et al. (2018).

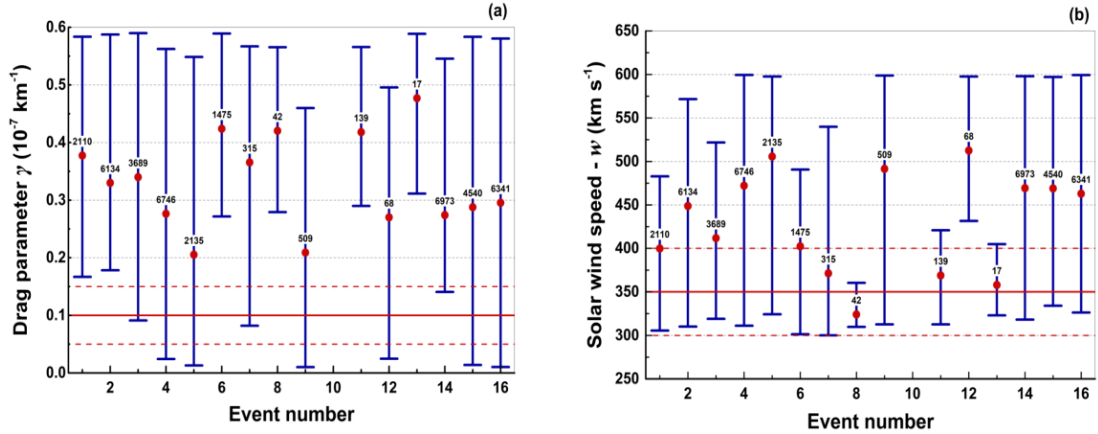

**Figure 3**: Optimal values for drag parameter, $\gamma$ (Figure 3a) and ambient solar wind speed, $w$ (Figure 3b), used as inputs in DBEM model to match the perfect ToA (within ±1 hour of actual observed ToA). The range of values found for each event is represented with the blue bars and the median values are marked with the red dots. The number above each red dot indicates the number of found results (values) in 50 000 runs that matched perfect ToA. Red solid and dotted lines denote mean and uncertainty range of $\gamma$ and $w$ values used in Dumbovic et al. (2018).

## 5. Discussion and Conclusions

Our work has a twofold objective. First, to improve the empirical Effective Acceleration Model (new version: EAMv3) used for the estimation of the shock arrival time (ToA) and the ICME speed upon arrival (SoA). Second, to investigate the propagation effects on CMEs/ICMEs using this EAMv3 and the ensemble version of the drag-based model (DBEM).

The improved version of EAMv3 model is based on: a) the conclusion that the calculation of the ICME effective acceleration or deceleration reduces uncertainties by using the minimum number of inputs (e.g. only LASCO linear speed – see Equation 3) and b) a new linear relation for the corrected speed of the ICME (Equation 6).

EAMv3 is based on the same sample of 214 ICMEs used in our previous work (Paper I) covering the years 1996-2009. The new EAMv3 has a MAE of 12.7 hours which implies an improvement of almost 6 hours in contrast to the first version (EAMv1) with a MAE of 18.6 hours (Table 1). This performance is comparable with the MAE of 10.4 hours from the study of Wold et al. (2018) where they also used a very large sample of 273 CMEs. At this point we should mention that a further improvement might be possible if the EAM model developed utilizing a sample of CMEs with their radial speeds (e.g. using multi-viewpoint observations) and not the projected ones from a single view point (e.g. LASCO). It would be interesting to examine how the radial speed calculated using multi-viewpoint observations affects the performance of the EAM model presented here. This will be studied in a future work. The validation and verification of EAMv3 model is performed with the same CME/ICME list of events, used in Mays et al. (2015) and Dumbovic et al. (2018). This list contains 16 CME/ICME events during the years 2013-2014. These CMEs/ICMEs events are fast events with a mean value of their initial speed equal to almost 1100 km s$^{-1}$. In particular, 9 CMEs has average speeds between 550 and 900 km s$^{-1}$ and 7 CMEs has speeds greater than 1000 km s$^{-1}$.

For the first time, to the best of our knowledge, a comparison between three very different models, i.e. an empirical data driven model (EAMv3), a drag-based model (DBEM) and an MHD analytical model (WSA-ENLIL+Cone model) is presented using the same sample and the same metrics. These metrics are the mean error (ME), the mean absolute error (MAE) and the root mean squared error (RMSE). All models overestimate the ToA with a ME of -9.67±3.43, -6.13±3.27 and -1.36±2.66 hours for DBEM, ENLIL and EAMv3 respectively. The MAE for these models are 14.31±2.18, 12.84±1.65 and 8.65±1.57 hours (Table 2).

The CME kinematic properties such as the POS speed and the angular width are suffering from projection effects (see e.g. Burkepile et al., 2004; Vrsnak, 2007; Paouris et al., 2020) that lead to underestimation of the true radial CME speed and overestimation of the angular width (Burkepile et al., 2004; Temmer et al., 2009). A method to obtain a CME speed close to the true radial speed is multi viewpoint reconstruction (Mierla et al., 2008, Temmer et al., 2009, Lee et al., 2015). Therefore, we use the mean values of 3D CME speeds, obtained utilizing the StereoCAT tool (see Mays et al., 2015). These speeds used also as input in the works of Mays et al. 2015 and Dumbovic et al. 2018. The new linear relation of the radial speed as a function of the LASCO linear speed (Equation 6), used by EAMv3 model, with a slope of almost 0.5 perhaps indicates a stronger deceleration in further distance from Sun (beyond 30 $R_{Sun}$), something which is not captured by the triangulation method for the sample of 16 ICMEs.

Wood et al. (2017), studied 28 CMEs associated with magnetic clouds during the period 2008-2012 and found that the deceleration occurs beyond 10 $R_{Sun}$ in most cases. Liu et al. (2013), also finds a similar result for a case study of three ICMEs. In particular, the phase of the rapid deceleration lasts about 5–10 hours and stops around 40–80 solar radii from the Sun for the examined cases. The very fast ICME of 7 March 2012 (peak speed of 2000 km s$^{-1}$) accelerates up to 10-15 solar radii and then decelerates rapidly out to about 50 solar radii. The moderate ICME of 12 December 2008 (700 km s$^{-1}$) accelerates with a lower rate up to 20 solar radii and then decreased out to 80-90 solar radii (see also Manchester et al., 2017).

So, we perform the following test to examine this scenario. We derive the distances where the initial CME speed ($u_0$) becomes equal to the corrected speed ($u_r$) as it is calculated by Equation 6. The input parameters for this process are the 3D CME speeds, under the assumption that these speeds are very close to the real ones, and the optimal acceleration (columns 5 and 6) of Table 3. In 11 cases it was possible to calculate these distances and we found a mean distance of 0.72 AU supporting our assumption. This result is not a surprising one. Gopalswamy et al. (2001), mentioned that the acceleration cessation distance is 0.76 AU and is in reasonable agreement with observations.

One could say that this result is biased by the basic assumption of EAM model that: the ICME interacts with the ambient solar wind with a constant "effective acceleration" from the Sun up to 1 AU, and this is correct, at least, up to a point. Nevertheless, it is still a very interesting result confirming also previous works (see Liu et al., 2013; Wood et al., 2017). Our assumption for deceleration in further distances, which are not actually captured due to the physical limitations in the FOV of the coronagraphs, is possible to be –finally– a fact and not just a hypothesis. Our work is in agreement with the results mentioned by Wood et al. (2017). This work proves our hypothesis as well as in 14 cases with CME speeds greater than 600 km s$^{-1}$ (fast CMEs as in our sample), calculated measurable deceleration in the interplanetary medium, after reaching their peak velocities. Many of these CMEs still exhibit some degree of deceleration even in distances of 1 AU.

We find that CMEs are decelerating more effectively in IP space than assumed. This is evident by the need to increase significantly (up to ≈ 3x) the drag parameter magnitude in drag-based models such DBEM. We reach this conclusion by using the known CME onset and arrival times to derive the optimal drag parameter values for perfect ToA (within ±1 hour of actual observed ToA). The mean value of the drag parameter for these 15 ICMEs is $0.33 \times 10^{-7}$ km$^{-1}$ while in the work of Dumbovic et al. (2018), it was $0.1 \times 10^{-7}$ km$^{-1}$ which is greater by a factor of ≈ 3. We are going to investigate this very interesting subject also utilizing measurements from the SECCHI/HI instruments tracking ICMEs the whole way from Sun up to Earth in future work.

We realize that the results could be different for a different sample, e.g. if the sample of CMEs consists of slower CMEs or for a different period of solar cycle. Nevertheless, these results may be very important for space weather forecasts, as the most important CMEs for geomagnetic effects and Solar Energetic Particles (SEPs) are fast Halo CMEs (see e.g. Zhang et al., 2007; Gopalswamy, 2009; Dumbovic et al, 2015; Papaioannou et al., 2016; 2018). Our future work will try to quantify these errors by utilizing as many as possible ICMEs from various published lists e.g. list in Paper I and Vourlidas et al. (2017) providing useful information for each class of CMEs as a function of their initial linear speed.

**Acknowledgments**

We are grateful to the providers of the solar data used in this work. The coronal mass ejection data are taken from the SOHO/LASCO CME list (http://cdaw.gsfc.nasa.gov/CME_list/). This CME catalog is generated and maintained at the CDAW Data Center by NASA and The Catholic University of America in cooperation with the Naval Research Laboratory. SOHO is a project of international cooperation between ESA and NASA.

E.P. supported by the project "PROTEAS II" (MIS 5002515), which is implemented under the Action "Reinforcement of the Research and Innovation Infrastructure", funded by the Operational Programme "Competitiveness, Entrepreneurship and Innovation" (NSRF 2014–2020) and co-financed by Greece and the European Union (European Regional Development Fund). A.A. and G.B. partially supported by the same project "PROTEAS II".

J.C. and M.D. acknowledge funding from the EU H2020 grant agreement No. 824135 (SOLARNET) and support by the Croatian Science Foundation under the project 7549 (MSOC).

A.V. is supported by NASA grants NNX17AC47G and 80NSSC19K0069.

A.P. and A.A. acknowledge the support through the ESA Contract No. 4000120480/NL/LF/hh ''Solar Energetic Particle (SEP) Advanced Warning System (SAWS)" and the TRACER project (http://members.noa.gr/atpapaio/tracer/), funded by the National Observatory of Athens (NOA) (Project ID: 5063).

E.P., M.D. and M.L.M. acknowledge International Space Science Institute (ISSI) team "Understanding Our Capabilities in Observing and Modeling Coronal Mass Ejections", led by C. Verbeke and M. Mierla.

**Conflict of interest:** The authors have no conflicts of interest to declare that are relevant to the content of this article.

**ORCID IDs:**
Evangelos Paouris Orcid: http://orcid.org/0000-0002-8387-5202
Jasa Calogovic Orcid: https://orcid.org/0000-0002-4066-726X
Mateja Dumbovic Orcid: https://orcid.org/0000-0002-8680-8267
M. Leila Mays Orcid: https://orcid.org/0000-0001-9177-8405
Angelos Vourlidas Orcid: https://orcid.org/0000-0002-8164-5948
Athanasios Papaioannou Orcid: https://orcid.org/0000-0002-9479-8644
Anastasios Anastasiadis Orcid: https://orcid.org/0000-0002-5162-8821
Georgios Balasis Orcid: https://orcid.org/0000-0001-7342-0557

**References**

Burkepile, J. T., Hundhausen, A. J., Stanger, A. L., St. Cyr, O. C., & Seiden, J. A.: 2004, *J. Geophys. Res.* 109, 3103. DOI: 10.1029/2003JA010149

Brueckner, G.E., Howard, R.A., Koomen, M.J., Korendyke, C.M., Michels, D.J., Moses, J.D., Socker, D.G., Dere, K.P., Lamy, P.L., Llebaria, A., Bout, M.V., Schwenn, R., Simnett, G.M., Bedford D.K., & Eyles, C.J.: 1995, *Solar Phys.* 162, 357-402. DOI: 10.1007/BF00733434

Colaninno, R.C., Vourlidas, A., Wu, C.C.: 2013, *J. Geophys. Res.* 118, 6866–6879. DOI: 10.1002/2013JA019205

Corona-Romero, P., Gonzalez-Esparza, J.A., Aguilar-Rodriguez, E., De-la Luz, V., Mejia-Ambriz, J.C.: 2015, *Solar Phys.* 290, 1–16. DOI: 10.1007/s11207-015-0683-2

Corona-Romero, P., Gonzalez-Esparza, J.A., Perez-Alanis, C.A., Aguilar-Rodriguez, E., de-la Luz, V., Mejia-Ambriz, J.C.: 2017, *Space Weather* 15, 464–483. DOI: 10.1002/2016SW001489

Domingo, V., Fleck, B. and Poland, A.I.: 1995, *Solar Phys.* 162, 1–37. DOI: 10.1007/BF00733425

Dumbovic, M., Devos, A., Vrsnak, B., Sudar, D., Rodriguez, L., Ruzdjak, D., Leer, K., Vennerstrom, S. and Veronig, A.: 2015, *Solar Phys.* 290, 579–612. DOI: 10.1007/s11207-014-0613-8

Dumbovic, M., Calogovic, J., Vrsnak, B., Temmer, M., Mays, M. L., Veronig, A., and Piantschitsch, I.: 2018, *Astrophys. J.* 854, 180. DOI: 10.3847/1538-4357/aaaa66

Fry, C. D., Sun, W., Deehr, C. S., Dryer, M., Smith, Z., Akasofu, S.-I., Tokumaru, M. and Kojima, M.: 2001, *J. Geophys. Res.* 106, 20,985–21,002. DOI: 10.1029/2000JA000220

Gopalswamy, N., Lara, A., Yashiro, S., Kaiser, M.L., Howard, R.A.: 2001, *J. Geophys. Res.* 106, (29), 207-218. DOI: 10.1029/2001JA000177

Gopalswamy, N., Lara, A., Manoharan, P. K., and Howard, R. A.: 2005, *Adv. Space Res.* 36, 2289–2294. DOI: 10.1016/j.asr.2004.07.014.

Gopalswamy, N.: 2009, In: Tsuda, T., Fujii, R., Shibata, K., Geller, M.A. (Eds.), Climate and Weather of the Sun-Earth System (CAWSES) Selected Papers from the 2007 Kyoto Symposium. Terrapub, Tokyo, pp. 77–120.

Gosling, J.T.: 1993, *J. Geophys. Res.* 98, 18937. DOI: 10.1029/93JA01896.

Hanslmeier, A.: 2010, in Gopalswamy N., Hasan S., Ambastha A. (eds) Heliophysical Processes. Astrophysics and Space Science Proceedings. Springer, Berlin, Heidelberg, DOI: 10.1007/978-3-642-11341-3_14

Hess, P., Zhang, J.: 2015, *Astrophys. J.* 812, 144. DOI: 10.1088/0004-637X/812/2/144

Kay, C., Mays, M. L., and Verbeke, C.: 2020, *Space Weather*, 18. DOI: 10.1029/2019SW002382

Kilpua, E.K.J., Mierla, M., Rodriguez, L., Zhukov, A.N., Srivastava, N., West, M.J.: 2012, *Solar Phys.* 279, 477–496. DOI: 10.1007/s11207-012-0005-x

Leblanc, Y., Dulk, G. A., Vourlidas, A., and Bougeret, J.L.: 2001, *J. Geophys. Res.* 106 (A11), 25301–25312. DOI: 10.1029/2000JA000260

Lee, H., Moon, Y.J., Na, H., Jang, S. and Lee, J.O.: 2015, *J. Geophys. Res. Space Physics*, 120, 10, 237-10, 249. DOI: 10.1002/2015JA021118

Liu, J., Ye, Y., Shen, C., Wang, Y. and Erdelyi, R.: 2018, *Astrophys. J.* 855, 109. DOI: 10.3847/1538-4357/aaae69

Liu, Y.D., Luhmann, J.G., Lugaz, N., Möstl, C., Davies, J.A., Bale, S.D., Lin, R.P.: 2013, *Astrophys. J.* 769, 45. DOI: 10.1088/0004-637X/769/1/45

Makela, P., Gopalswamy, N., Yashiro, S.: 2016, *Space Weather* 14, 368–378. DOI: 10.1002/2015SW001335

Manchester, W., Kilpua, E.K.J., Liu, Y.D., Lugaz, N., Riley, P. et al.: 2017, *Space Sci. Rev.* 212, 1159–1219. DOI: 10.1007/s11214-017-0394-0

Mays, M. L., Taktakishvili, A., Pulkkinen, A., MacNeice, P. J., Rastatter, L., Odstrcil, D., Jian, L.K., Richardson, I.G., LaSota, J.A., Zheng, Y. and Kuznetsova, M.M.: 2015, *Solar Phys.* 290, 1775–1814. DOI: 10.1007/s11207-015-0692-1

McKenna-Lawlor, S. M. P., Dryer, M., Kartalev, M. D., Smith, Z., Fry, C. D., Sun, W., Deehr, C.S., Kecskemety, K., Kudela, K.: 2006, *J. Geophys. Res.* 111, A11103. DOI: 10.1029/2005JA011162

Millward, G., Biesecker, D., Pizzo, V., de Koning, C.A.: 2013, *Space Weather* 11, 57–68. DOI: 10.1002/swe.20024

Mierla, M., Davila, J., Thompson, W., Inhester, B., Srivastava, N., Kramar, M., St. Cyr, O.C., Stenborg, G. and Howard, R.A., 2008, *Solar Phys.* 252, 385–396. DOI: 10.1007/s11207-008-9267-8

Mostl, C., Amla, K., Hall, J.R., Liewer, P.C., De Jong, E.M., Colaninno, R.C., et al.: 2014, *Astrophys. J.* 787, 119. DOI: 10.1088/0004-637X/787/2/119

Mostl, C., Isavnin, A., Boakes, P.D., Kilpua, E.K.J., Davies, J.A. et al.: 2017, *Space Weather* 15, 955–970. DOI: 10.1002/2017SW001614

Napoletano, G., Forte, R., Moro, D.D., Pietropaolo, E., Giovannelli, L., Berrilli, F.: 2018, *J. Space Weather Space Clim.* 8, A11. DOI: 10.1051/swsc/2018003

Paouris, E., and Mavromichalaki, H.: 2017a, *Solar Phys.* 292, 30. DOI: 10.1007/s11207-017-1050-2

Paouris, E., and Mavromichalaki, H.: 2017b, *Solar Phys.* 292, 180. DOI: 10.1007/s11207-017-1212-2

Paouris, E., Vourlidas, A., Papaioannou, A., and Anastasiadis, A.: 2020, "Assessing the projection correction of Coronal Mass Ejection speeds on Time-of-Arrival prediction performance using the Effective Acceleration Model", *Space Weather*, *accepted*. DOI: 10.1029/2020SW002617

Papaioannou, A., Sandberg, I., Anastasiadis, A., Kouloumvakos, A., Georgoulis, M.K., Tziotziou, K., Tsiropoula, G., Jiggens, P., Hilgers, A.: 2016, *J. Space Weather Space Clim.* 6, A42, DOI: 10.1051/swsc/2016035

Papaioannou, A., Anastasiadis, A., Sandberg, I., and Jiggens, P.: 2018, *J. Space Weather Space Clim.* 8, A37. DOI: 10.1051/swsc/2018024

Papaioannou, A., Belov, A., Abunina, M., Eroshenko, E., Abunin, A., Anastasiadis, A., Patsourakos, S., Mavromichalaki, H.: 2020, *Astrophys. J.* 890, 101. DOI: 10.3847/1538-4357/ab6bd1

Pomoell, J., and Poedts, S.: 2018, *J. Space Weather Space Clim.* 8, A35. DOI: 10.1051/swsc/2018020

Riley, P., Mays, M.L., Andries, J., Amerstorfer, T., Biesecker, D. et al.: 2018, *Space Weather* 16, 1245–1260. DOI: 10.1029/2018SW001962

Rollett, T., Mostl, C., Isavnin, A., Davies, J.A., Kubicka, M., Amerstorfer, U.V., and Harrison, R.A.: 2016, *Astrophys. J.* 824, 131. DOI: 10.3847/0004-637X/824/2/131

Schwenn, R., Dal Lago, A., Huttunen, E., Gonzalez, W.D.: 2005, *Ann. Geophys.* 23, 1033–1059. DOI: 10.5194/angeo-23-1033-2005

Shi, T., Wang, Y., Wan, L., Cheng, X., Ding, M., Zhang, J.: 2015, *Astrophys. J.* 806, 271. DOI: 10.1088/0004-637X/806/2/271

Shiota, D., and Kataoka, R.: 2016, *Space Weather* 14, 56–75. DOI: 10.1002/2015SW001308

Sudar, D., Vrsnak, B., Dumbovic, M.: 2016, *Mon. Not. R. Astron. Soc.* 456, 1542–1548. DOI: 10.1093/mnras/stv2782

Szabo, A.: 1994, NASA Wind Satellite. In: Pelton J., Allahdadi F. (eds) Handbook of Cosmic Hazards and Planetary Defense. Springer, Cham. DOI: 10.1007/978-3-319-03952-7_13

Temmer, M., Preiss, S. and Veronig, A.M.: 2009, *Solar Phys.* 256, 183-199. DOI: 10.1007/s11207-009-9336-7

Vourlidas, A., Patsourakos, S. and Savani, N. P.: 2019, Philosophical Transactions of the Royal Society A: Mathematical, Physical and Engineering Sciences, 377. DOI: 10.1098/rsta.2018.0096

Vourlidas, A., Balmaceda, L. A., Stenborg, G., and Dal Lago, A.: 2017, *Astrophys. J.* 838, 141. DOI: 10.3847/1538-4357/aa67f0

Vrsnak, B., Zic, T., Vrbanec, D., Temmer, M., Rollett, T., Mostl, C. et al.: 2013, *Solar Phys.* 285, 295–315. DOI: 10.1007/s11207-012-0035-4

Vrsnak, B., Sudar, D., Ruzdjak, D., and Zic, T.: 2007, *Astron. Astrophys.* 469, 339. DOI: 10.1051/0004-6361:20077175

Wold, A.M., Mays, M.L., Taktakishvili, A., Jian, L.K., Odstrcil, D., MacNeice, P.: 2018, *J. Space Weather Space Clim.* 8, A17. DOI: 10.1051/swsc/2018005

Wood, B.E., Wu, C.C., Lepping, R.P., Nieves-Chinchilla, T., Howard, R.A., Linton, M.G., Socker, D.G.: 2017, *Astrophys. J. Suppl.* 229, 29. DOI: 10.3847/1538-4365/229/2/29

Zhang, J., Richardson, I.G., Webb, D.F., Gopalswamy, N., Huttunen, E., Kasper, J.C. et al.: 2007, *J. Geophys. Res.* 112, A10102. DOI: 10.1029/2007JA012321